\documentclass{iaus}
\usepackage{graphics,epsfig}

  \checkfont{eurm10}
  \iffontfound
    \IfFileExists{upmath.sty}
      {\typeout{^^JFound AMS Euler Roman fonts on the system,
                   using the 'upmath' package.^^J}%
       \usepackage{upmath}}
      {\typeout{^^JFound AMS Euler Roman fonts on the system, but you
                   dont seem to have the}%
       \typeout{'upmath' package installed. iaus.cls can take advantage
                 of these fonts,^^Jif you use 'upmath' package.^^J}%
      }
  \else
  \fi

  \checkfont{msam10}
  \iffontfound
    \IfFileExists{amssymb.sty}
      {\typeout{^^JFound AMS Symbol fonts on the system, using the
                'amssymb' package.^^J}%
       \usepackage{amssymb}%
       \let\le=\leqslant  \let\leq=\leqslant
         
      }{}
  \fi

  \IfFileExists{amsbsy.sty}
    {\typeout{^^JFound the 'amsbsy' package on the system, using it.^^J}%
     \usepackage{amsbsy}}
    {\providecommand\boldsymbol[1]{\mbox{\boldmath $##1$}}}

\newsavebox{\astrutbox}
\sbox{\astrutbox}{\rule[-5pt]{0pt}{20pt}}

\title[Outskirts of Galaxy Clusters: intense life in the suburbs]
      {The Stellar Populations of Low-redshift Clusters}

\author[Craig Harrison {\it et al.\/}]%
{Craig Harrison$^1$,\,Matthew Colless$^2$,\,Warrick J. Couch$^3$\break
\and B. A. Peterson$^1$}

\affiliation{$^1$Mount Stromlo Observatory, The Australian National
  University, Cotter Road, Weston, ACT 2611, Australia email:
  harrison@mso.anu.edu.au,\,peterson@mso.anu.edu.au\\[\affilskip] 
$^2$Anglo-Australian Observatory, PO Box 296, Epping, NSW 1710,
  Australia \break email: colless@aao.gov.au\\[\affilskip]
$^3$School of Physics, University of New South Wales, Sydney NSW 2052,
  Australia \break email: wjc@phys.unsw.edu.au}  
\pubyear{2004}
\volume{195}
\pagerange{1--8}
\date{?? and in revised form ??}
\jname{Outskirts of Galaxy Clusters: intense life in the suburbs}
\editors{A. Diaferio, ed.}
\begin{document}

\maketitle

\begin{abstract}
We present some preliminary results from an on-going study of the
evolution of stellar populations in rich clusters of galaxies. Our
baseline sample contains core line-strength measurements from 183
galaxies with $b_{\rm J} \le 19.5$ from four clusters with
$\bar{z}\sim0.04$, against which observations of higher-redshift
clusters can be compared.  Using predictions from stellar population
models to compare with our measured line strengths we can derive
\textit{relative} luminosity-weighted mean ages and metallicities for
the stellar populations in each of our clusters. It must be stressed
that these ages and metallicities are only accurate when used in an
relative sense as the stellar population models, due to differing
abundance ratios used in the models compared to those observed in
elliptical galaxies, provide inaccurate absolute ages and
metallicities. We also investigate the
Mg\textit{b}$^\prime$-$\sigma_0$ and H$\beta_G^\prime$-$\sigma_0$
scaling relations. We find that Mg\textit{b}$^\prime$ is correlated
with $\sigma_0$, the likely explanation being that larger galaxies are
better at retaining their heavier elements due to their larger
potentials. H$\beta_G^\prime$, on the other hand, we find to be
anti-correlated with $\sigma_0$. This result implies that the stellar
populations in larger galaxies are older than in smaller galaxies.
\end{abstract}

\section{Introduction and Observations}\label{sec:intro}

The formation and evolution of clusters of galaxies and of their
early-type galaxy population is a topic of active debate. Studying the
integrated light from stellar populations can help to discriminate
between models of giant elliptical galaxy formation: the monolithic
collapse model (\cite[Eggen et al., 1962]{eggen62}) in which giant
elliptical galaxies form rapidly in a process that can essentially be
considered a single collapse and the hierarchical merging model
(\cite{searle78}) in which giant elliptical galaxies are built up over
a long timescale by distinct merger events. If it was found that
early-types were largely coeval in clusters then this would favour the
monolithic collapse model, whereas a large spread in ages would favour
hierarchical merging.

Increasing a galaxy's age by a factor of two while decreasing it's
metallicity by a factor of three results in a spectrum almost
indistinguishable from the original (\cite{worthey94}). This
age/metallicity degeneracy can be partially broken by comparing
age-sensitive spectroscopic indices with metallicity-sensitive
spectroscopic indices. By plotting two such indices against each other
and comparing the results to stellar population models
(e.g. \cite{worthey94}; \cite{thomas03}) we can determine the mean
relative luminosity-weighted age and metallicity of a given stellar
population.

Previous work in this area has concentrated on individual clusters
such as Coma (\cite{jorgensen99}; \cite{moore02}) and Fornax
(\cite{kuntschner00}) with sometimes conflicting results coming from
the same cluster, e.g. \cite[Gonzalez et al. (1993)]{gonzalez93} and
\cite[Moore et al. (2002)]{moore02}. Coma and Fornax represent
the extremes of the cluster richness scale and this project will study
clusters that fill in this range. We will study, in detail, the
stellar populations in several low-redshift ($z\sim0.04$) clusters
from the core out into the surrounding structures. These results will
then be used as a comparison for observations of a range of
higher-redshift clusters ($0.3\leq z \leq 0.55$), allowing us to draw
conclusions about the formation and evolution of the stellar
populations in rich clusters.

Line-strength measurements, redshifts and velocity dispersions were
obtained for 183 galaxies from four clusters over three nights in
April 2002 with the 2dF system on the 3.9m AAT. The sample was
selected from galaxies with $b_{\rm J}\le19.5$, within a $2^{\rm o}$
diameter FOV centred on each of the clusters and within a $3\sigma$
redshift range centred on the cluster redshift. The 300B grating
($\sim9$\AA\, FWHM) was used to obtain a wide spectral range covering
as many indices as possible, while the 1200V grating ($\sim2$\AA\,
FWHM) was used to obtain higher resolution spectra yielding precise
velocity dispersions.

\begin{figure}
  \epsfig{file=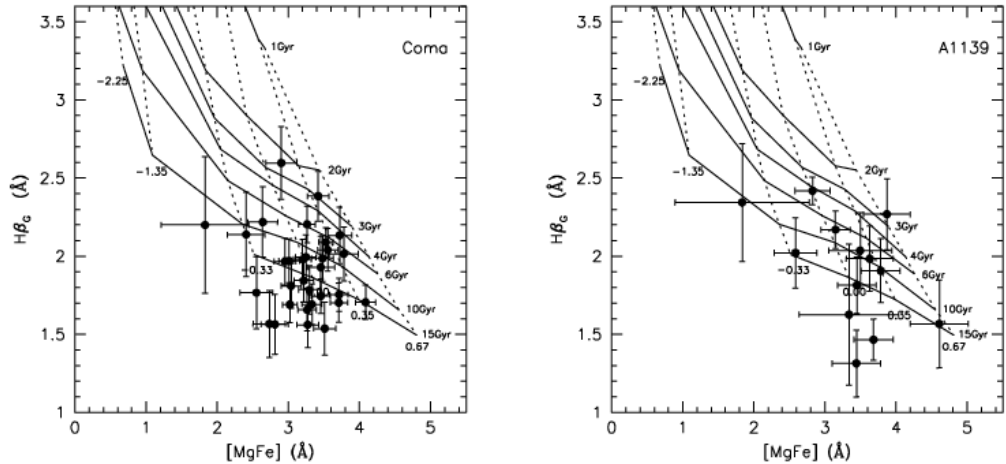,width=\textwidth}
  \caption{H$\beta_G$ equivalent width versus [MgFe] equivalent
    width. \cite[Thomas et al. (2003)]{thomas03} stellar population model grids
    have been overlaid on data from the Coma cluster (\textit{left})
    and Abell 1139 (\textit{right}).}
  \label{fig:thomas} 
\end{figure}

\section{Ages and Metallicities}\label{fig:ages}

Comparison of the measured equivalent widths of H$\beta_G$ and [MgFe]
for two of our clusters, Coma and A1139, and predictions from the
\cite[Thomas et al. (2003)]{thomas03} stellar population models is
shown in Fig. \ref{fig:thomas}. Using these models we obtain relative
ages and metallicities for our stellar populations. We find that Coma
cluster galaxies have a mean age of $9.3\pm 3.8$ Gyr with a mean
metallicity of $0.15\pm 0.23$ dex relative to solar. A1139, we find,
has a mean age of $10.3\pm 3.4$ Gyr with a mean metallicity of
$0.12\pm 0.41$ dex relative to solar. Refinements still need to be
made to the data set and the other clusters need to be analysed before
any conclusions can be drawn. The age/metallicity degeneracy is quite
evident from these plots. Also, there seems to be some disagreement
between our observations and the models. The data points with low
H$\beta_G$ equivalent widths could be star-forming galaxies, which we
wish to exclude from this part of the analysis. By examining each
galaxy's spectra for emission lines indicative of a star-forming
galaxy and by measuring H$\alpha$ equivalent widths we can eliminate
these galaxies from our sample.
  
\section{Scaling Relations}\label{sec:scaling}

\begin{figure}
  \epsfig{file=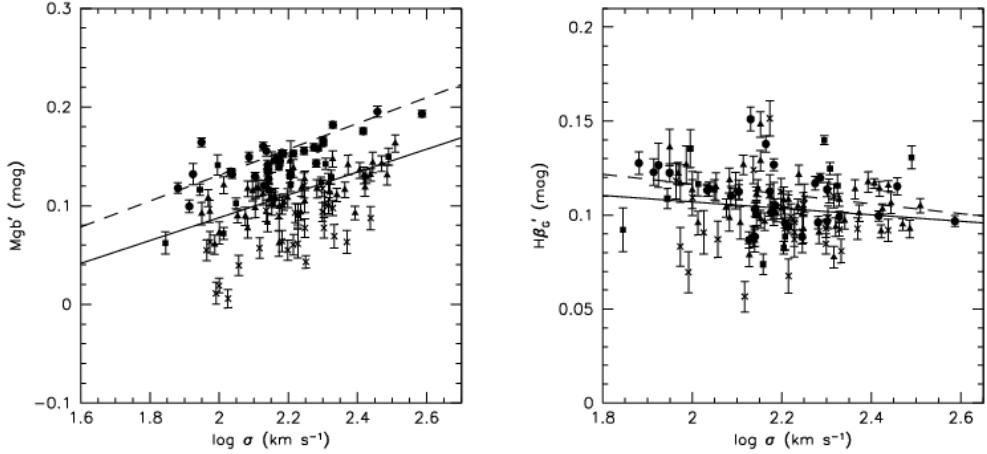,width=\textwidth}
  \caption{Scaling relations in low-redshift clusters. (\textit{left})
    The Mg$b$' versus $\sigma_0$ relation; the solid line is our fit
    to the data and the dashed line is that from \cite[Colless et
    al. (1999)]{colless99}. (\textit{right}) The H$\beta_G$' versus
    $\sigma_0$ relation; the solid line is our fit to the data and the
    dashed line is that from \cite[Kunstchner
    (2000)]{kuntschner00}. The symbols are the same for both figures:
    circles represent galaxies from Coma, squares from A1139, triangles
    from A3558 and crosses from A0930.}
  \label{fig:scaling} 
\end{figure}

In Fig. \ref{fig:scaling} we show the Mg\textit{b}$^\prime$ (the dash
denoting that the index value has been converted from Angstroms to
magnitudes) and H$\beta_G^\prime$ versus central velocity dispersion
($\sigma_0$) scaling relations for all four of our clusters. The solid
line in both plots are our robust fit to the data from all four
clusters while the dotted lines are fits taken from the literature.

We find that H$\beta_G^\prime$ is anti-correlated with $\sigma_0$ and
find good agreement between our fit of H$\beta_G^\prime=-0.017
\,\log\sigma_0 + 0.144$ and that from \cite[Kunstchner
(2000)]{kuntschner00} of H$\beta_G^\prime=-0.026\,\log\sigma_0 +
0.169$. The likely implication is that more massive galaxies are
older relative to less massive galaxies.


Mg\textit{b}$^\prime$, however, we find is correlated with
$\sigma_0$. The usual explanation for this correlation is that larger
ellipticals have a deeper potential and hence find it easier to hold
on to heavier elements produced in supernova explosions than do
smaller ellipticals. We obtain a fit of Mg\textit{b}$^\prime=0.116\,
\log\sigma_0 + 0.144$ while the fit found by \cite[Colless et
al. (1999)]{colless99} is Mg\textit{b}$^\prime=0.131\,\log\sigma_0 +
0.131$. The slopes agree quite well but there seems to be an offset
between our fit and \cite[Colless et al. (1999)]{colless99}. We also
note an offset in Mg\textit{b} between our
clusters. Since previous results have found no significant change in
the zero-point of this relation (e.g. \cite[Colless et
al., 1999]{colless99}) we suspect this may be a data reduction issue,
but it may possibly indicate cluster to cluster variations, contrary
to previous results.

\end{document}